\documentclass[floatfix,twocolumn,amsmath,amssymb,aps,prl,showpacs,letter,superscriptaddress]{revtex4}
\usepackage{latexsym,epsfig,bm,times,psfrag,subfigure}

\usepackage{color}
\usepackage{ulem}

\newcommand{\bs}[1]{\boldsymbol{#1}}

\newcommand{\tso}{Tb$_{2}$Sn$_{2}$O$_{7}$}
\newcommand{\hto}{Ho$_{2}$Ti$_{2}$O$_{7}$}

\newcommand{\dto}{Dy$_{2}$Ti$_{2}$O$_{7}$}

\newcommand{\tto}{Tb$_{2}$Ti$_{2}$O$_{7}$}

\begin{document}
\title{Soft dipolar spin ice physics from an effective Hamiltonian description
  and the ordered phase of the frustrated Tb$_{2}$Sn$_{2}$O$_{7}$ pyrochlore magnet}
\title{Soft dipolar spin ice physics and the ordered phase of the frustrated Tb$_{2}$Sn$_{2}$O$_{7}$ pyrochlore magnet}
\author{Paul A. McClarty} 
\affiliation{Department of Physics and Astronomy, University of Waterloo, Waterloo, ON, N2L 3G1, Canada.}
\author{Pawel Stasiak}
\affiliation{Department of Physics and Astronomy, University of Waterloo, Waterloo, ON, N2L 3G1, Canada.}
\affiliation{Department of Mathematics, University of Reading, Whiteknights, PO Box 220, Reading RG6 6AX, UK.}
\author{Michel J. P. Gingras} 
\affiliation{Department of Physics and Astronomy, University of Waterloo, Waterloo, ON, N2L 3G1, Canada.}
\affiliation{Canadian Institute for Advanced Research, 180 Dundas Street West, Suite 1400, Toronto, ON, M5G 1Z8, Canada.}
\date{\today} 
\begin{abstract}
From a microscopic model for the pyrochlore antiferromagnet \tso, including the crystal
field Hamiltonian and interactions between the angular momenta, we compute an effective
pseudospin-1/2 Hamiltonian $H_{\rm eff}$ that incorporates perturbatively in
the effective interactions the
effect of excited crystal field levels. We obtain the semiclassical ground states of $H_{\rm eff}$
and find a region of parameter space with a two-in/two-out spin ice configuration on each
tetrahedron with ordering wavevector $\mathbf{q}=0$ and with spins canted away from the local
Ising axes as found in \tso. This ground state can also be obtained 
from a dipolar spin ice model in which
the Ising constraint is softened. 
Monte Carlo simulations on the latter model reveal a region of the phase
diagram with spin ice-like freezing and another with a transition into \tso-type long range order. We
comment on the differences between \tso\ and the perplexing spin liquid \tto.
\end{abstract}


\maketitle


In recent years, much effort has been devoted to the exploration of
geometrically frustrated magnetic systems \cite{Diep,Balents,SpinIce,RMP}. 
Among the many such materials, the spin ices \hto\ and \dto\ 
\cite{SpinIce} are remarkable for at least two reasons. 
Firstly, they exhibit a low temperature strongly correlated state 
with residual magnetic entropy and no long range order. 
Secondly, the mechanism leading to effectively
geometrically frustrated interactions in spin ices and the equilibrium
low-temperature properties that follow from them have been definitively understood 
\cite{SpinIce,RMP,DSI,self-screening,Castelnovo}. 
Progress on spin ices has been made quantitatively
through the dipolar spin ice model (DSIM) \cite{SpinIce,DSI} and, 
conceptually, via an understanding of the self-screening of the dipolar interactions
\cite{self-screening,Castelnovo}
and a description of the low-energy excitations as deconfined magnetic charges
\cite{Castelnovo,Jaubert}.

That progress on spin ices has been made so steadily can be traced partly
to the fact that their magnetic moments can be treated as classical and
Ising-like. This is due to the fortuitous smallness of
 the inter-ion interactions compared to the
single ion crystal field anisotropy gap. 
But this is also good fortune for those interested
in the problem of quantum fluctuations in strongly correlated
geometrically frustrated systems for, in spin ices, 
one can explore quantum effects by perturbing away from the
Ising limit by reducing the anisotropy gap. Moreover, this is not merely an
academic problem for it directly pertains to the \tso\ (TSO) \cite{TSO3}
and \tto\ (TTO) \cite{TTO} compounds which both exhibit rich, complex and poorly understood
behaviors.

TSO has a $0.87$ K transition to a long range ordered state \cite{TSO3,TSO1}
where the magnetic
moments obey the same (two-in/two-out) ``ice rules'' that hold in spin ice materials \cite{SpinIce}. 
However, the ordered moments in TSO 
are slightly canted away from the Ising ``in/out'' directions 
while displaying fluctuations that persist down to the lowest temperature considered. 
TTO, in contrast, despite a Curie-Weiss temperature 
$\theta_{\rm CW}\sim -14$ K set by the magnetic interactions, 
fails to develop long range order
down to at least $50$ mK \cite{TTO}, 
making it a rare example of a spin liquid in three dimensions \cite{Balents}. 
Despite numerous experiments aimed at exposing the
essential physics at play in TSO 
\cite{TSO3,TSO1,Mirebeau_CF,TSO4,TSO_Pressure} and TTO \cite{TTO,TTO2,TTOHysteresis,Hamaguchi},
no microcopic theory has yet been able to explain the behavior of these two materials.

Just as perturbative calculations that treat
the de Broglie wavelength as a small parameter allow one to 
describe quantum
corrections to the properties of simple liquids
(e.g. argon), in the same spirit, we examine in this paper 
the effects of weak quantum mechanical corrections to the DSIM $-$
a classical spin liquid of sorts \cite{Balents}. 
As a key milestone, we consider TSO. 
By displaying long range order, TSO
is  particularly amenable to conventional
experimental probes of its physics. 
This allows us to benchmark our calculations against known experimental results.

 Starting from a microscopic model for TSO, we
derive a low energy effective Hamiltonian, $H_{\rm eff}$, which incorporates
virtual crystal field excitations (VCFEs) about the DSIM \cite{TTOHeff}. 
We compute the semiclassical ground states of $H_{\rm eff}$ and find
 three phases in the vicinity of the Ising
limit $-$ one of which has the same magnetic structure as found in TSO \cite{TSO3}. 
We therefore provide a realistic microscopic
explanation for the magnetic order of TSO. 
To explore the extent to which TSO-like order is generic among pyrochlore
oxides with competing exchange, dipoles and single ion anisotropy,
we consider a toy model, $H_{\rm m}$, with 
 explicit tunable single ion anisotropy.
While we find that $H_{\rm m}$ admits a TSO-like phase over
a wide range of parameters, we observe that
a spin ice-like state with dynamically inhibited LRO  persists
close to the Ising limit. 
By exploring the phase diagram over the entire
parameter space, we  identify 
a  fanning out of several different phases
upon tuning away from the classical Heisenberg antiferromagnet spin 
liquid \cite{MoessnerChalker}. 
Finally, on the basis of
this work, we offer a novel perspective on how TTO relates to TSO.


{\it Effective Hamiltonian} $-$ The simplest microscopic
Hamiltonian consistent with the properties of Tb$^{3+}$ ions in TSO and
with an antiferromagnetic $\theta_{\rm CW}$ is
$ H = H_{\rm cf} + V$
where $H_{\rm cf}$ is the crystal field Hamiltonian with parameters taken from
Ref.~\cite{Mirebeau_CF}. The interaction term, $V=H_{\rm ex}+ H_{\rm dd}$, is
the sum of the nearest neighbor isotropic exchange  $H_{\rm ex} = \mathcal{J}_{{\rm ex}}\sum_{\langle i,j\rangle}
\mathbf{J}_{i}\cdot\mathbf{J}_{j}$ between angular momenta $\mathbf{J}_{i}$
(${\rm J}=6$) on site $i$ and dipole-dipole interactions   $H_{\rm dd} =
\mathcal{D}r_{{\rm nn}}^{3}\sum_{{i>j}}
[\mathbf{J}_{i}\cdot\mathbf{J}_{j} -
3(\mathbf{J}_{i}\cdot\mathbf{\hat{R}}_{ij})(\mathbf{J}_{j}\cdot\mathbf{\hat{R}}_{ij})]|\mathbf{R}_{ij}|^{-3}$.
The dipole coupling $\mathcal{D}=\mu_{0}(g\mu_{B})^{2}/4\pi r_{\rm nn}^{3} = 0.029$ K where the
 Land\'{e} factor of Tb$^{3+}$ is $g=3/2$ and the exchange has been estimated to be
$\mathcal{J}_{\rm ex}\sim 0.08$ K \cite{Mirebeau_CF}. $\mathbf{R}_{ij}=\mathbf{R}_{i}-\mathbf{R}_{j}$ where
$\mathbf{R}_{i}$ is the position of the magnetic ion on site $i$. The nearest
 neighbor distance  $r_{{\rm nn}}=3.69\mbox{\AA} = a\sqrt{2}/4$ \cite{TSO3} 
and $a$ is the edge length of the cubic unit cell.
Throughout this paper, 
the long range dipolar interactions are handled via the Ewald method \cite{Enjalran}.


Diagonalizing $H_{\rm cf}$ gives a spectrum of $2{\rm J}+1=13$ states with an Ising-like ground state
doublet separated from the first excited doublet 
by a gap $\Delta\sim 13.8$ K \cite{Mirebeau_CF}. 
Because the interactions, $V$, are much smaller than $\Delta$, we treat $V$ as a
perturbation. Using degenerate perturbation theory,
we derive an effective Hamiltonian, $H_{\rm eff}$, acting in the
space $\mathfrak{M}$ spanned by the ground doublet on each site and which
includes, to leading order in $V^2/\Delta$, the effect of admixing of the
excited crystal field levels into the low energy space (i.e. VCFEs) \cite{TTOHeff}. 
Because the low energy space is two-dimensional on
each site, $H_{\rm eff}$ can be rendered in the form of a pseudospin-1/2
($\bs{S}_{\rm eff}$) Hamiltonian \cite{TTOHeff}. 
In the limit $V/\Delta\rightarrow 0$,  $H_{\rm eff} \sim O(V)$ is the projection of
$V$ onto the ground doublet which gives the dipolar spin ice Ising model
(DSIM) \cite{DSI,TTOHeff,SpinIceReview}.
Terms to order $V^2/\Delta$ deform $H_{\rm eff}$ away from the Ising limit.

\begin{figure}
\begin{center}
\includegraphics[width=0.5\textwidth]{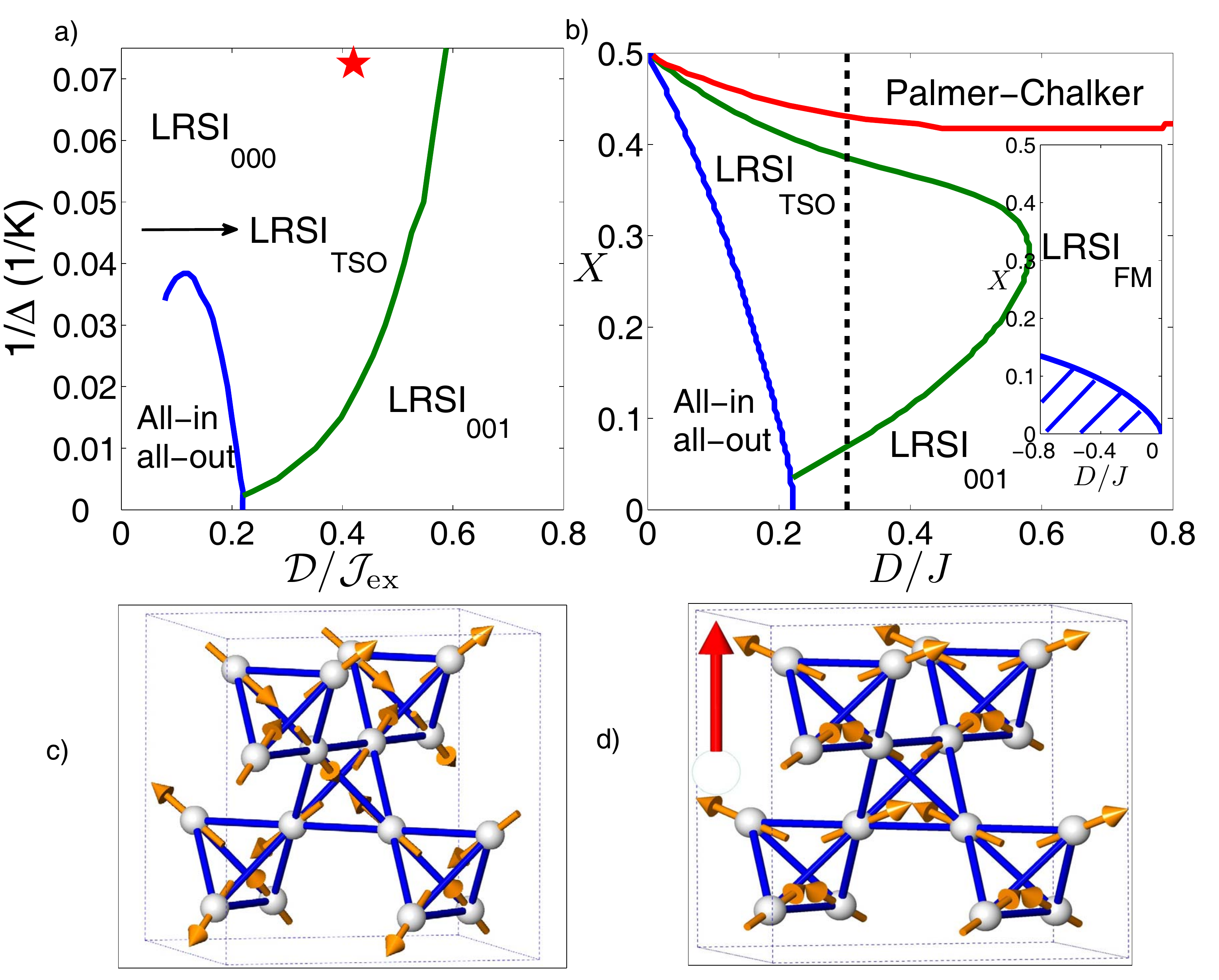}
\caption{\label{fig:groundstates} (color online). (a) Semiclassical ground
  state of $H_{\rm eff}$ on a single cubic unit cell as a function of
  $1/\Delta$ and $\mathcal{J}_{\rm ex}/\mathcal{D}$. The star indicates the
  estimated microscopic parameters of TSO. (b) Ground states of the
  classical spin model $H_{\rm m}$ as a function of couplings $D/J$ and
  the anisotropy $X$. Ground states are shown for antiferromagnetic $J$ and an Ising
  anisotropy. 
Note the similar topology of the phase diagrams in (a) and (b) close to
$D/J,{\mathcal{D}}/{\mathcal{J}_{\rm ex}} \approx 0.2$ and
small anisotropy (small $1/\Delta,X$).
The inset shows the ground states for ferromagnetic $J$ for $0<X<1$.
 The vertical dashed line indicates the line
  along which Monte Carlo simulations have been carried out. The lower
  panels show the LRSI$_{\rm TSO}$ (c) and LRSI$_{001}$ (d) spin
  configurations on a single cubic unit cell.
 The vertical (red) arrow in (d)
  shows the $[0,0,1]$ direction of the bulk net moment.}
\end{center}
\end{figure}

{\it $H_{\rm eff}$ ground states} $-$ To examine the effects of excited crystal field 
levels on the spin correlations, we
determine the semiclassical ground states of $H_{\rm eff}$ 
on a cubic unit cell with periodic boundary
conditions by replacing the pseudospins, 
${\bs S}_{\rm eff}({\mathbf R}_i)$, 
by classical spins of fixed length $1/2$.
This is akin to finding a semiclassical N\'{e}el
ordered phase as the leading description of the broken symmetry phase in a
quantum Heisenberg antiferromagnet. 
The ground state phase diagram as  the anisotropy gap $\Delta$ and
$\mathcal{D}/\mathcal{J}_{\rm ex}$ are varied is shown in panel (a) of
Fig.~\ref{fig:groundstates}. 
For $\Delta=13.8$ K corresponding to TSO, the ground state for weakly antiferromagnetic
exchange $\mathcal{J}_{{\rm ex}}<0.050$ K ($\mathcal{D}/\mathcal{J}_{\rm ex}>0.58$) 
has a $\mathbf{q}=001$ ordering wavevector with zero bulk magnetization and the 
two-in/two-out spin ice
rule satisfied by the local $[111]$ (Ising) 
components of ${\bs S}_{\rm eff}$ on each tetrahedron. 
Three body interactions in H$_{\rm eff}$ cause the 
${\bs S}_{\rm eff}(\mathbf{R}_{i})$ to cant away from the local Ising directions. 
The ``uncanted'' variant of this
$\mathbf{q}=001$ long range ordered spin ice (LRSI$_{001}$) state is one of the two
ground states of the DSIM for $1/\Delta=0$ \cite{DSI} (see Fig.~\ref{fig:groundstates}).
For $\mathcal{J}_{{\rm ex}}>0.050$ K 
($\mathcal{D}/\mathcal{J}_{\rm ex}  < 0.58$)
and $\Delta=13.8$ K, instead of the
all-in/all-out state observed in the DSIM \cite{DSI} (which
appears here only for $1/\Delta\lesssim 0.04$ K$^{-1}$), 
we find that the ground state is an ordered
ice state with ordering wavevector $\mathbf{q}=0$ and with the ice rule
satisfied on each tetrahedron (LRSI$_{000}$). Quantum fluctuations away from
these classical ground states are currently being investigated and will be
reported elsewhere. The all-in/all-out to LRSI$_{000}$ phase boundary, 
computed for fixed $\mathcal{D}$, has a
maximum for small $\mathcal{D}/\mathcal{J}_{\rm ex}$ which indicates
that terms to order $V^2/\Delta$ in perturbation theory are not adequate to
account for the phase boundary for too small $\Delta$.


The ground states spin configurations determined from $H_{\rm eff}$ are the semiclassical
expectation values of the pseudospins, ${\bs S}_{\rm eff}$. However, the
physical observables are, rather, the 
expectation values of the angular momenta 
$\langle {{\rm J}}^{\tilde \alpha}\rangle$ for 
${\tilde \alpha}={\tilde x}, {\tilde y}, {\tilde z}$. 
We therefore compute the $\langle {\rm J}^{\tilde{\alpha}}\rangle$ 
from the $S_{\rm eff}^{\tilde \alpha}$ expectation values. 
We do this perturbatively in the interactions
$V$ finding
\begin{equation}    \langle {{\rm J}}^{\tilde \alpha}\rangle = 
\langle \mathcal{P} {{\rm J}}^{\tilde \alpha}\mathcal{P}\rangle +
\langle \mathcal{P}V\mathcal{Q}{\rm J}^{\tilde{\alpha}}\mathcal{P}\rangle +
\langle \mathcal{P} {{\rm J}}^{\tilde \alpha}\mathcal{Q}V\mathcal{P}\rangle + \ldots  
\label{eqn:correction}
\end{equation}
for the expectation values of 
$\langle {{\rm J}}^{\tilde \alpha}\rangle$ within the low energy
 subspace where  $\mathcal{Q}=\sum_{|\psi\rangle\in
 \mathfrak{M}}|\psi\rangle\langle\psi|/(E_{g}-E_{|\psi\rangle})$ and $E_{g}$
 is the ground state energy of the unperturbed ground crystal field doublet.
 The tilde over the components of $\langle {\rm J}^{\tilde\alpha}\rangle$ 
indicates that the components are 
taken in the local  ${\tilde x}-{\tilde y}-{\tilde z}$ 
frame on each sublattice \cite{Enjalran}. The expansion in operators on the right hand
 side of Eq.~(\ref{eqn:correction}) can be rendered in the form of pseudospin-1/2
 operators with coefficients that are computed numerically.

We are mainly interested in the $\mathbf{q}=0$ ordered ice state (LRSI$_{000}$). 
We compute $H_{\rm eff}$ on a cubic unit cell and consider a state vector, within
the pseudospin $\mathfrak{M}$ space, corresponding to the LRSI$_{000}$ with
canted ${\bs S}_{\rm eff}({\mathbf{R}}_i)$
that minimizes the classical energy. The state vector
corresponding to such semiclassical ground state is a direct product of canted
pseudospins-1/2 states on each site of the cubic cell. The expectation value of
${{\rm J}}^{\tilde z}$, $\langle {{\rm J}}^{\tilde z}\rangle$, is computed from
Eq.~(\ref{eqn:correction}) on each of the $16$ sites as a function of $\mathcal{J}_{\rm ex}$. 
As one would expect from Eq.~(\ref{eqn:correction})
the magnitude of $\langle{{\rm J}}^{\tilde z}\rangle$ changes as $\mathcal{J}_{\rm ex}$ varies. 
However, the relative signs of ${{\rm J}}^{\tilde z}$ on each of the sites are {\it preserved},
so the identification of the $\mathbf{q}=0$ ordered spin ice state is
{\it unaltered} by the operator correction of Eq.~(\ref{eqn:correction}). Next,
we find the $\langle {{\rm J}}^{\tilde x}\rangle$ and 
$\langle {{\rm  J}}^{\tilde y}\rangle$ expectation 
values on each lattice site for which the only
contributions come from the terms linear in $V$ in
Eq.~(\ref{eqn:correction}). The result of this calculation is that the
$\langle {{\rm J}}^{\tilde \alpha}\rangle$ ground state configuration is the
same magnetic structure of canted magnetic moments as observed in TSO \cite{TSO3} 
$-$ a structure that we henceforth refer to as LRSI$_{\rm TSO}$.
Note that it is crucial that the ordered moments be described via
the observables  $\langle {{\rm J}}^{\tilde \alpha} \rangle$
in Eq.~(\ref{eqn:correction}), as sole consideration of the 
expectation values of the ${\bs S}_{\rm eff}(\mathbf{R}_i)$ 
does not even give  qualitatively the correct canting 
direction and disagrees with the canting experimentally observed in TSO \cite{TSO3}.


{\it Soft dipolar spin ice model} $-$ The essential physics of $H_{\rm eff}$ is
that VCFEs lead to (i) a softening of the
otherwise Ising spins and (ii) a new (LRSI$_{000}$) ordered spin ice state not
present in the DSIM \cite{DSI}. To explore the effects of
the softened Ising constraint and the extent to which the
appearance of a LRSI$_{000}$ state is generic in dipolar pyrochlore systems
with softened $[111]$ Ising anisotropy, we investigate a toy model,
$H_{\rm m}$, with an explicitly anisotropic $g$-tensor.

We take $H_{\rm m} \equiv H_{\rm ex} + H_{\rm dd}$ as used in the
 microscopic model $H$ discussed above where the  $\mathbf{J}_{i}$
are taken to be Heisenberg spins. For this model, we denote
 the exchange and dipole couplings by $J$ and $D$ respectively. 
We incorporate 
 the anisotropy by introducing spins-1/2 ${\bs S}$ with components defined via
$({{\rm J}}^{\tilde x}, {{\rm J}}^{ \tilde  y}, {{\rm J}}^{\tilde z}) \rightarrow (g_{\perp}
{{ S}}^{\tilde x}, g_{\perp} {{ S}}^{\tilde y}, g_{\parallel} {{ S}}^{\tilde z})$
 with $g_{\parallel}> g_{\perp}$ to achieve an Ising-like anisotropy. 
Whereas,  in $H_{\rm eff}$, the spin softness is generated through VCFEs, in $H_{\rm m}$
 we include the effects of this physics  at the outset. 
We parameterize the anisotropy $\mathbf{g}\equiv
 (g_{\perp},g_{\perp},g_{\parallel})$ by the parameter $X$ such that
 $\mathbf{g}= \eta(X,X,1-X)$ with
 $\eta$ chosen so that $\mathbf{g}$ is a unit vector. 
When $X=0$, (i.e. $g_{\perp}=0$), we obtain
 a model that couples only the Ising components of the ${\bs S}$
 vectors. 
This model for $J=0$ was studied in Ref.~\cite{DipolarGround} and a
 related model with $D=0$, $X=1/2$ and ferromagnetic $J$, but with a single ion anisotropy
 $-\Gamma(S^{z}_{i})^{2}$, has been investigated \cite{SoftSpinIce}.

Panel (b) of Figure~\ref{fig:groundstates} displays
the semiclassical ground states of $H_{\rm m}$ as $X$
and $D/J$ are varied. The ground states are computed on a single cubic unit
cell using zero temperature Monte Carlo. The main figure shows the ground
states for $0\leq X<0.5$ (Ising anisotropy) and antiferromagnetic $J>0$. 
The point $(D/J=0,X=0.5)$ corresponds to the classical
Heisenberg antiferromagnet spin liquid on 
the pyrochlore lattice \cite{MoessnerChalker}, from which
four long range ordered phases fan out.
For small $D/J$, the ground state is an
all-in/all-out (AIAO) state with ordering wavevector $\mathbf{q}=0$ with no
canting of the spins away from the Ising directions. 
For large $D/J$, one finds
LRSI$_{001}$ states with ordering wavevector $\mathbf{q}=001$ with the ice rule satisfied on
each tetrahedron. The AIAO state and the LRSI$_{001}$ state are both ground states of the DSIM and
the boundary between them for $X=0$ is consistent with earlier findings
\cite{DSI}. We find that the canting angle away from the Ising
directions varies smoothly away from $0$ (for $X=0$) in the LRSI$_{001}$ state as the parameters vary. In the
isotropic ($X=1/2$, Heisenberg) limit, we recover the Palmer-Chalker ground state
\cite{PalmerChalker}. We find that this state remains the ground state even away
from the isotropic limit with spins lying in the local ${\tilde x}{\tilde y}$ planes. 
At the center of the figure is a dome in which the ground state is
a LRSI$_{\rm TSO}$ state with $D/J$ and $X$ dependent canting angles. The
inset shows the ground states for ferromagnetic exchange ($J<0$). The LRSI$_{001}$ ground
state (for $X=0$) extends to finite $X$ (dashed region),
whereupon it gives way to a $\mathbf{q}=0$ two-in/two-out ordered state that 
we label LRSI$_{\rm  FM}$. The distinction between LRSI$_{\rm TSO}$ and LRSI$_{\rm FM}$ is that, in the former
(latter), the net moment on each tetrahedron is smaller (larger) than the moment $4\mu/\sqrt{3}$ 
of a $\mathbf{q}=0$ two-in/two-out
state with moments $\mathbf{\mu}$ constrained to lie along the $\langle 111\rangle$ Ising directions. 
The labelling of the
regions in the figure is independent of the canting angle away from the $\langle
  111\rangle$ directions. The canting angle (not indicated) increases as $X$
  increases with discontinuities at the phase boundaries. Within the
  LRSI$_{\rm TSO}$ phase, the general
 trend in the canting angle is for it to decrease as
  $D/J$ increases.
  
\begin{figure}
\begin{center}
\includegraphics[width=0.5\textwidth]{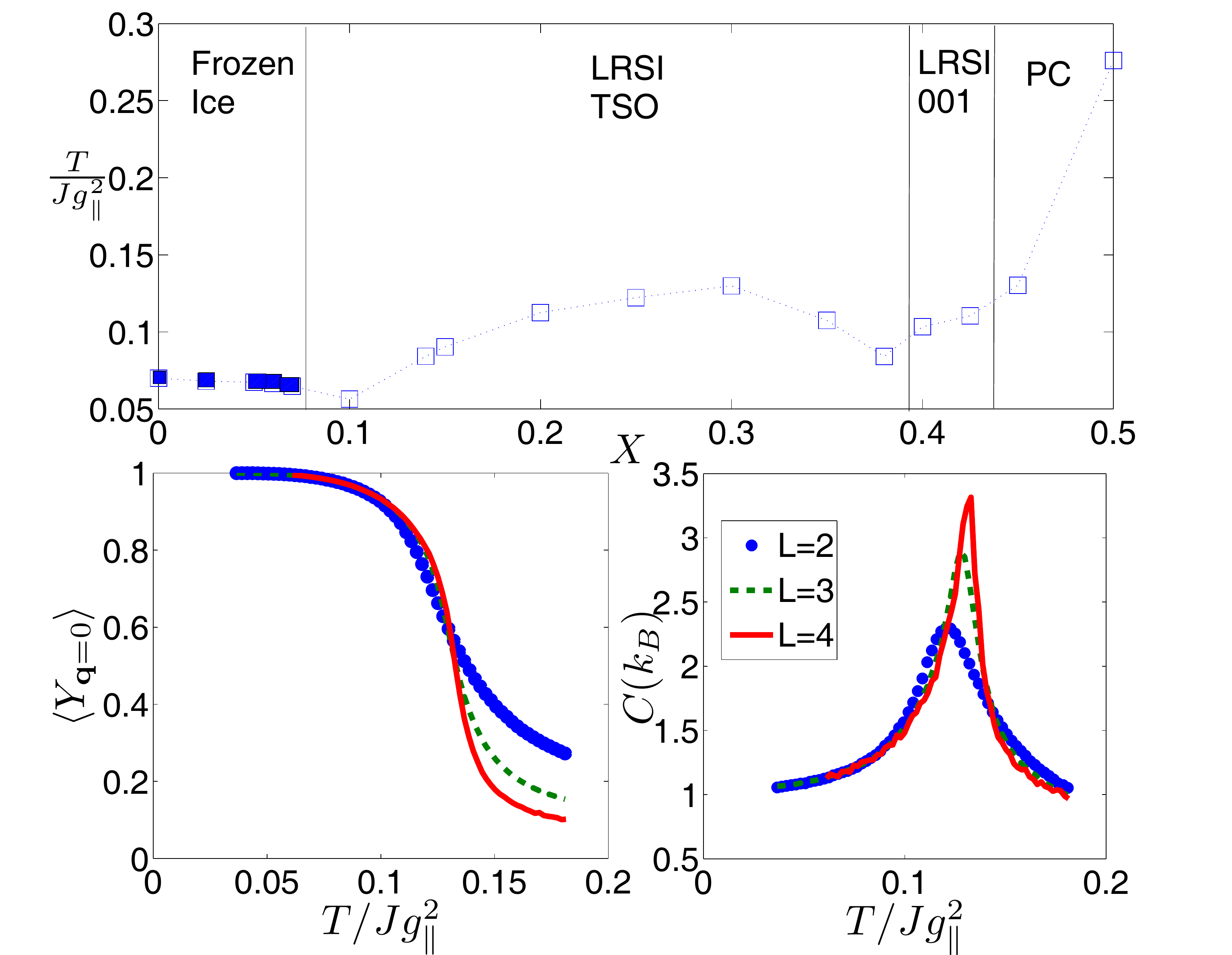}
\caption{\label{fig:Simulations} (color online). Selected Monte Carlo simulation results on $H_{\rm
    m}$. Simulations have been carried out along the line $D/J=0.3$. The upper panel shows the
    freezing temperatures (for $X<0.07$) and transition temperatures ($X>0.07$) determined from the
    heat capacity peak for system size $L=3$. The ground states as a function of $X$ are
    indicated. The lower left hand panel shows, for $X=0.25$, the increase of the $\mathbf{q}=0$
    ordered ice order parameter $Y_{\mathbf{q}=0}$ at the temperature is lowered for $L=2,3,4$. The
    lower right hand panel shows the heat capacity for $X=0.25$ and $L=2,3,4$.}
\end{center}
\end{figure}

We have explored the finite temperature phase diagram along the line
$D/J=0.3$ using Monte Carlo (MC) simulations with parallel tempering. We chose this
line because it cuts through three sets of ground states as $X$ is varied (see panel (b)
of Fig.~\ref{fig:groundstates}). For Ising interactions, $X=0$, upon lowering the temperature, the
conventional paramagnet  {\it freezes} into a spin ice state 
characterized by a two-in/two-out constraint on each tetrahedron but {\rm without} conventional LRO. While
one would expect some residual entropy, in this case, it does not coincide with the Pauling entropy of 
spin ice
because the spins are classical 3-component spins with a magnetic specific heat per spin of $k_{B}$ at $T=0$
(see lower right panel of Fig. 2). 
Of particular interest, this freezing is observed at finite $X$ {\it up to} the
boundary between LRSI$_{001}$ and LRSI$_{\rm TSO}$.
The freezing temperatures are indicated in the top panel of Fig.~\ref{fig:Simulations} for
$X \lesssim 0.07$. 
For $X \gtrsim 0.07$ and $D/J=0.3$, the simulations equilibrate easily with strong evidence of phase
transitions into long-range ordered phases corresponding to the ground states
of Fig.~\ref{fig:groundstates}. 
The transition temperatures from the heat capacity peak are plotted in the top
panel of Fig.~\ref{fig:Simulations}. 
The lower panels of Fig.~\ref{fig:Simulations} 
shows for $X=0.25$, within the LRSI$_{\rm TSO}$ dome, 
the order parameter $Y_{\mathbf{q}=0}$ 
(an order parameter for the $\mathbf{q}=0$ ordered ice \cite{Yq0})
and the specific heat  for different system sizes measured in
$L$ $-$ the length of the cubic simulation cell in units of the cubic unit
cell edge.
Since the simulations reveal a freezing transition 
for $X \lesssim 0.07$ and LRO at finite temperature in the
(reentrant) LRSI$_{001}$ region around $X=0.4$ (see top panel of Fig.~\ref{fig:Simulations}), 
there exists a crossover between a freezing transition 
($X\lesssim X^*$) and a phase transition to LRO  ($X\gtrsim X^*$) within the LRSI$_{001}$
region of Fig.~\ref{fig:groundstates}b. We have run a series of simulations
for $L=2,3$ and $4$ finding that this boundary lies near $X^*\approx 0.35$, 
largely independently of $D/J$ as far as we can tell. We always observe the onset of
finite temperature LRO within the LRSI$_{\rm TSO}$ dome in Fig.~\ref{fig:groundstates}b.

{\it Materials context} $-$ Given the considerable
microscopic similarities between TSO and TTO, one wonders whether $H_{\rm m}$ might
help shed some light on the low temperature collective paramagnetism of TTO. Since
the anisotropy gap in TTO is larger than in TSO, one might consider
TTO to have the larger Ising anisotropy (smaller $X$)  and to lie in the frozen canted spin
ice regime of $H_{\rm m}$. Interestingly, TTO does exhibit hysteresis and a frequency
dependent a.c. susceptibility \cite{TTOHysteresis,Hamaguchi} below about $350$ mK which is consistent with a
slowing down of the dynamics. However, this glassiness has been suggested to
come from only a fraction of the moments \cite{TTOHysteresis}, the rest exhibiting fast spin dynamics
down to the lowest observed temperatures consistent with the lack of LRO \cite{TTO}.
 On the other hand, low temperature ($T\lesssim 500$ mK) spin dynamics
does not distinguish the two materials for they are also observed
in TSO in spite of the LRO \cite{TSO4}. 
More critically, we note that the mechanism
underlying the persistent spin dynamics in these materials as in other
pyrochlores with LRO and gapped collective excitations 
(e.g. Gd$_{2}$Sn$_{2}$O$_{7}$
\cite{GSO-muSR}) remains an important open problem \cite{RMP,GSO-muSR}. 
Since $H_{\rm m}$ captures the LRO in TSO which remains dynamical,
we cannot rule out
the possibility that dynamics may also occur in TTO at $T\lesssim 350$ mK 
{\it alongside} spin ice-like freezing as we find in $H_{\rm m}$. 
Further investigations will be necessary to settle this matter.

{\it Conclusion} $-$
We have shown that a low energy effective
Hamiltonian $H_{\rm eff}$ for \tso\ derived from a microscopic model including  crystal
field, antiferromagnetic exchange and dipolar interactions exhibits
semiclassical ground states coinciding with the magnetic structure of
\tso\ (LRSI$_{\rm TSO}$ states) \cite{TSO3}. We used this result to motivate 
the study of a classical
spin model $H_{\rm m}$ referred to as a soft dipolar spin ice model (SDSIM). This model also exhibits
a finite temperature transition into LRSI$_{\rm TSO}$ states over a large portion 
of its phase diagram indicating
that it captures the essential physics obtained from the microscopic
model. Therefore, we are confident that we have correctly identified the
predominant physics leading to the $\mathbf{q}=0$ spin ice ordered phase of
\tso\ and the concurrent spin canting away from the Ising directions. A corollary of our results is that
spin-lattice couplings are not necessary to account for the long range ordered phase
of \tso. More generally, the SDSIM reveals the competing and important effects of single 
ion anisotropy, exchange
and dipolar interactions in rare earth pyrochlore oxides. We hope that
our toy model $H_{\rm m}$ will serve as a bridge between these materials
and ultimately help to unravel the fascinating phenomena at play in the \tto\
spin liquid material.

We thank B. Gaulin, P. Holdsworth and J. Ruff for useful
discussions. This research was funded by the NSERC of Canada and the Canada
Research Chair program (M. G., Tier I). We acknowledge the use of
computational resources from SHARCNET.


\begin{thebibliography}{99}
\bibitem{Yq0} This order parameter is defined as follows: we introduce vectors
  $I^{\alpha}_{u}$ with components $I_{1}=(1,1,-1,-1)$, $I_{2}=(1,-1,-1,1)$
  and $I_{3}=(-1,1,1,-1)$. These are the three ways in which Ising components
  of spins in spin ice can be arranged on a single tetrahedron. Note that
  $\mathbf{I}_{u}\cdot\mathbf{I}_{v} = 4\delta_{uv}$. The order parameter
  $Y_{\mathbf{q}=0}$ is $(1/N)|\sum_{u}\sum_{i,a} \sigma_{i}^{a}I_{u}^{a}$ 
where
  $\sigma_{i}^{a}=\pm 1$ is the sign of the Ising component of the spin on
  sublattice $a$, fcc vector $i$ and $N$ is the number of spins. $\langle
  Y_{\mathbf{q}=0}\rangle$ denotes the thermally averaged order parameter.
\end{thebibliography}
\end{document}